\documentclass{article}

\PassOptionsToPackage{numbers,sort&compress}{natbib}
\usepackage[preprint]{neurips_2025}
\usepackage[utf8]{inputenc}
\usepackage[T1]{fontenc}
\usepackage{hyperref}
\usepackage{url}
\usepackage{xurl}
\usepackage{booktabs}
\usepackage{amsmath,amsfonts,amssymb}
\usepackage{nicefrac}
\usepackage{microtype}
\usepackage{xcolor}
\usepackage{graphicx}
\usepackage{tabularx}
\usepackage{array}
\usepackage{placeins}
\usepackage[inline,shortlabels]{enumitem}

\hypersetup{
  hidelinks,
  pdftitle={From Subjective Judgments to Auditable Standards: Protocol-Guided AI Auditing of Website Redundancy},
  pdfauthor={Ge Kong; Yongtong Cao},
  pdfsubject={Protocol-guided and calibration-gated AI auditing of website redundancy},
  pdfkeywords={website redundancy, AI auditing, auditable standards, construct validity, evidence grounding, score refusal}
}

\setlist{leftmargin=1.35em,itemsep=0.2em,topsep=0.3em}
\newcommand{\cora}{\textsc{Cora}}
\newcommand{\aabs}{\textsc{A-Abs}}
\newcommand{\cord}{\textsc{C-Ord}}
\newcommand{\withhold}{\texttt{WITHHOLD\_AUTOMATED\_SCORES}}
\newcolumntype{Y}{>{\raggedright\arraybackslash}X}
\newcolumntype{P}[1]{>{\raggedright\arraybackslash}p{#1}}

\title{From Subjective Judgments to Auditable Standards:\\
Protocol-Guided AI Auditing of Website Redundancy}
\author{
  Ge Kong \\
  Beihang University \\
  \texttt{gekong@buaa.edu.cn}
  \And
  Yongtong Cao \\
  Beijing Institute of Technology \\
  \texttt{yongtong.cao@bit.edu.cn}
}

\begin{document}
\maketitle

\begin{abstract}
Website redundancy does not have a single fixed meaning. The same repeated
element may distract during one task and provide backup during another. We
introduce \cora{} (Counterfactual, Observable Redundancy Audit), which measures
repetition \emph{load}, normal-use \emph{tax}, and failure-domain recovery
\emph{reserve} separately. Each run retains screenshots, stable element
identities, and task traces. A versioned vision-language model proposes the
annotations. Typed validation and release checks then determine whether a
calibrated dimension can be reported; failed or malformed outputs stay in the
fixed denominator.

On a transparent mechanistic testbed, the factorized \cora{} representation
separated reserve from normal-use tax and predicted perturbed success more
accurately than scalar-load baselines. The model studies then showed why
repeatability is not enough: two small local
vision-language models produced recurring outputs, but neither instrument met
all release requirements. \cora{} therefore withheld automated scores from
both instruments while retaining the raw responses and failure records.
Separate checker fixtures confirmed that the typed validator and hardened
release gates implement their specifications; these tests do not establish
semantic grounding or accuracy on production sites.

Taken together, the results position \cora{} as an auditable candidate
procedure for the controlled benchmark studied here rather than a general
standard. Human agreement, AI-versus-human accuracy, and validation on
independent production sites remain open empirical questions.
\end{abstract}

\noindent\textbf{Keywords:} AI-assisted usability evaluation; website
redundancy; construct validity; counterfactual evaluation; multimodal language
models; evidence grounding; selective score release.

\section{Introduction}

Website redundancy is easy to name but difficult to measure consistently. Its
meaning changes with the user, task, failure scenario, and form of repetition.
Prior work likewise finds that redundancy depends on the user and on how it is
represented \cite{reddy2020redundancy}. Evaluators using the same usability
method on the same interface can identify markedly different problem sets
\cite{hertzum2001evaluator}. We therefore specify the measurement procedure
before asking a model for a judgment. Human agreement and AI-versus-human
accuracy are outside the present experiments.

A frozen model, prompt, and browser configuration makes an audit repeatable,
but repeatability alone does not make the output valid. The same configuration
can consistently follow a bad output contract or miss the intended construct.
\cora{} treats the model as a versioned measurement instrument: it checks the
response against visible page evidence and controlled changes before reporting
a score. Validity concerns the meaning and use of a score-based inference even
when that inference is deterministic
\cite{cronbach1955construct,messick1995validity}. Computational measurement can
fail when a theoretical construct and its operationalization diverge
\cite{jacobs2021measurement}.

We measure three effects separately. \emph{Load} records repeated visual,
information, or interaction structure. \emph{Tax} is the work paid during
normal use, while \emph{reserve} counts cues, landmarks, or routes that remain
useful after a failure. A confirmation step can raise tax without adding
reserve; an alternate route adds reserve only if it survives a different
failure domain. A single ``clutter'' score cannot represent these distinctions.

\cora{} places a replaceable vision-language model (VLM) in an auditable
measurement pipeline. The model proposes semantic annotations and typed
citations. Independent checks determine whether a score can be reported. Before
execution, the audit contract fixes the task, rendered page, viewport, browser,
model revision, prompt, and success condition. The run stores screenshots, a
stable element catalog, and task traces; deterministic checks measure routes and
task states. The release rule applies only to dimensions covered by controlled
low/medium/high interventions. If any requirement fails, the result is
\withhold{}.

The model studies show why this separation matters. Generation settings,
prompts, thresholds, and metric concepts were locked before the corresponding
outputs, and documented conservative corrections were later applied to the
immutable-output analyses. Qwen recovered a small amount of repetition ordering
after a complete output-contract change but failed grounding and coverage.
SmolVLM2 echoed observation traces instead of returning the requested schema.
Both instruments were withheld, and their records show different reasons for
failure.

This paper makes four contributions.
\begin{enumerate}
    \item We define redundancy in terms of what is repeated (visual,
    information, and interaction \emph{load}), what ordinary use pays for it
    (planned and realized \emph{tax}), and what remains useful after a failure
    (cue, landmark, and route \emph{reserve}).
    \item We specify a conjunctive release gate. Parsing, evidence count,
    repetition ordering, grounding, target, qualified replay, and
    non-degeneracy checks must all pass before a controlled dimension is
    released. Uncontrolled reserve and clarity fields remain ineligible.
    \item We evaluate this qualification workflow with a transparent
    mechanistic testbed and four full-size seed replications, a 77-call
    discovery study, a 336-call calibration repair, and a 210-call second-model
    replication. These studies separate repeatability, discrimination,
    grounding, and safe refusal across model configurations.
    \item We test the checker independently of model quality using a
    504-citation typed grounding suite, a 672-string end-to-end fault suite, and
    a 336-string evidence-cardinality suite.
\end{enumerate}

The experiments test whether one frozen, falsifiable protocol keeps automated
judgments inspectable when the model changes and refuses release when its checks
fail.

\section{Related work}

\subsection{Appearance, complexity, and redundancy}

Interface appearance can shape perceived usability before interaction.
Tractinsky et al. studied the ``what is beautiful is usable'' effect
\cite{tractinsky2000beautiful}; Moshagen and Thielsch decomposed visual
aesthetics into distinct facets \cite{moshagen2010facets}; and Gritz et al.
related layout-level visual complexity to learning-oriented search outcomes
\cite{gritz2025visualcomplexity}. Visual clutter has also been operationalized
with image measures that correlate with visual-search performance in complex
imagery \cite{rosenholtz2007clutter}. Web studies treat perceived complexity as
a possible marker of cognitive load \cite{harper2009visualcomplexity} and show
that complexity and prototypicality shape rapid aesthetic judgments
\cite{tuch2012firstimpression}. Screenshot models can estimate perceived
complexity and colorfulness, while user characteristics still explain part of
the rating variation \cite{reinecke2013firstimpressions}.

These results make visual complexity measurable, but do not make it synonymous
with redundancy. Reddy et al. treated redundancy as an explicit UI design
variable and showed that its effects depend on user and representation rather
than being uniformly beneficial \cite{reddy2020redundancy}. \cora{} adds a
failure-domain distinction: repetition, nominal cost, and recovery capacity are
separate measurements.

\subsection{Human evaluation and measurement quality}

Usability is not one interchangeable outcome. Reviews find substantial
variation in how studies operationalize usability \cite{hornbaek2006measuring},
and a meta-analysis reports generally weak associations among effectiveness,
efficiency, satisfaction, and subjective versus objective measures
\cite{hornbaek2007metaanalysis}. Agreement measures also make different
assumptions and require interpretation in light of the rating design
\cite{artstein2008intercoder,krippendorff2004reliability}. Together with the
evaluator effect, this literature supports explicit rubrics and measured
reliability. \cora{} contributes an inspectable automated protocol; a future
human study must still measure inter-rater agreement rather than assume it.

\subsection{Automated UX evaluation and UI judges}

Automated usability evaluation predates current computer-use agents. Ivory and
Hearst organized earlier methods by degree of automation and described
automation as a way to augment evaluation \cite{ivory2001automating}. AIM later
exposed distinct computational GUI metrics, including visual clutter and visual
learnability, rather than collapsing them into one score
\cite{oulasvirta2018aim}. Recent systems and preprints extend this direction:
uxCUA learns interaction flows and usability scores \cite{gao2026uxcua};
Avenir-UX combines web interaction with structured UX instruments
\cite{tan2026avenir}; and UXCascade aggregates agent traces and issues for
practitioners \cite{holter2026uxcascade}.

Other work evaluates language and vision-language models as judges. Structured,
stepwise LLM evaluation can improve alignment with human judgments
\cite{liu2023geval}, while broader studies identify sensitivity to response
position, verbosity, self-produced answers, and other evaluator effects
\cite{zheng2023llmjudge,wang2024fairness,koo2024cognitive}.
UI-specific studies reach similarly qualified conclusions:
MLLM-as-a-UI-Judge reports dimension-dependent agreement with human preferences
\cite{luera2025mllmjudge}; WiserUI-Bench finds limited understanding of
behavior-changing UI differences \cite{jeon2025wiserui}; and agentic reward
systems incorporate static and interactive aesthetic judgments
\cite{xiao2025codeaesthetics}. This evidence shows why automated judgment is
worth testing and why the judge itself needs qualification. It does not
establish the bias profile of the two VLMs studied here.

GUI grounding is a separate capability from high-level judgment. SeeClick
treats element localization as a central bottleneck for visual GUI agents
\cite{cheng2024seeclick}, while ScreenAI uses dedicated element type and location
annotations for screen understanding \cite{baechler2024screenai}. Large
vision-language models can also produce object claims unsupported by an image
\cite{li2023objecthallucination}. These findings motivate separate checks for
structure, visible support, and the final judgment. They do not validate
\cora{}'s specific region-overlap threshold or evidence-count rule.

\subsection{Evidence attribution and verifiability}

Research on attributed generation separates a fluent answer from support for its
claims. Attribution evaluation asks whether an identifiable source supports a
claim \cite{rashkin2023attribution}; citation-aware generation benchmarks score
answer correctness separately from the recall and precision of citation support
\cite{gao2023citations}. \cora{} applies an analogous distinction to interface
measurement: evidence correctness and output-level coverage are distinct gate
conditions. Text-attribution studies, however, do not validate our region IoU
threshold or the requirement of 2--4 evidence items.

\subsection{Behavioral testing, validity, and abstention}

Behavioral test suites complement aggregate accuracy by checking known
capabilities and failure modes \cite{ribeiro2020checklist}. \cora{} applies this
principle to website measurement by testing whether controlled interventions
move the relevant output in the expected direction. Reproducibility programs and
model cards show the value of versioned artifacts and explicit reporting
\cite{pineau2021reproducibility,mitchell2019modelcards}; those practices improve
traceability but do not establish validity. Selective classification provides a
formal basis for abstention \cite{elyaniv2010selective,geifman2017selective},
usually in response to prediction confidence or coverage. Here, abstention
follows failures in parsing, discrimination, evidence, target behavior, replay,
or degeneracy. The resulting gate qualifies a complete measurement procedure;
it is not a confidence-based classifier.

\section{Measurement protocol}

\paragraph{Plain-language guide.}
\cora{} turns one vague redundancy judgment into three concrete questions: what
the page repeats, what that repetition costs during ordinary use, and what
remains useful when something fails. \emph{Load} records
repeated visual decoration, information, or interaction. \emph{Tax} records
ordinary-task costs such as extra actions. \emph{Reserve} records useful
structure that survives a matching failure, such as an alternate route when the
primary route is blocked. \emph{Evidence grounding} requires each model claim
to identify a specific visible element or region. Under a \emph{fixed
denominator}, all scheduled outputs remain in the calculation, including parse
failures, so retries cannot remove bad cases. The \emph{qualification gate}
releases a dimension only after independent checks pass, including a
\emph{non-degeneracy} check that prevents an instrument from passing by giving
the same answer to every page. The result is a reproducible record of the
proposal, its evidence, and the gate's release or refusal decision.

\subsection{Task-conditioned load, tax, and reserve}

Let a rendered interface be a state-action system $G=(S,A,T,O)$, where $S$ is
the interface state, $A$ the available actions, $T$ the transition function,
and $O$ the evidence visible under a fixed observation contract. A task $\tau$
provides an initial state, instruction, and machine-verifiable success states
$S^+_\tau$.

The load vector
\begin{equation}
L=(L_v,L_i,L_a)
\end{equation}
records repeated visual, information, and interaction structure. Planned tax
$C_{\mathrm{plan}}$ counts avoidable nominal action opportunities. Realized
path tax is
\begin{equation}
C_{\mathrm{real}}(G,\tau)=\mathbb{E}[|\pi|]-|\pi^*|,
\end{equation}
where $|\pi^*|$ is the family-specific minimum path. Structural reserve is
\begin{equation}
R=(R_{\mathrm{landmark}},R_{\mathrm{cue}},R_{\mathrm{route}}).
\end{equation}
For perturbation $\delta$, only matching failure domains contribute to
$R_\delta=w_\delta^\top R$. Robustness retention compares perturbed and nominal
success for the same composition:
\begin{equation}
Q(G,\tau,\delta)=
\frac{P(\mathrm{success}\mid G,\tau,\delta)}
{P(\mathrm{success}\mid G,\tau,\delta_0)}.
\end{equation}

\subsection{Typed evidence and fixed-denominator metrics}

Each candidate judgment cites observable catalog entries. A text citation must
name a valid ID and match the full normalized visible text. A region citation
must name a valid ID, leave its quote empty, and overlap the catalog box at
intersection-over-union at least 0.95. Text and region evidence are not
interchangeable. The output contract also requires 2--4 evidence items per
call; pooled grounding accuracy cannot substitute for output-level coverage.

Every scheduled output stays in the denominator. A parse failure is incorrect
for ordering and target metrics; it is never retried, repaired, or imputed.
Strict ordering requires all low $<$ medium $<$ high relations for the
manipulated repetition dimension. Studies C/D use a family-cluster bootstrap,
and their exact sign summaries use the six fixed family means. Study A reports
dispersion intervals over six overlapping leave-one-family-out folds, not a
population confidence interval. Conditional Kendall $\tau$ is reported only
where both vectors vary.

\subsection{Dimension-scoped release rule}

Let $\mathcal D_{\mathrm{pub}}$ be dimensions proposed for publication and
$\mathcal D_{\mathrm{ctrl}}$ dimensions with controlled interventions. Let $p$
be schema parse rate; let $c=1$ only when every schema-valid scheduled
stochastic and deterministic output contains 2--4 evidence items; let $d_{fs}$
be strict repetition ordering for family $f$ and seed $s$; and let $g$, $t$,
$r$, and $m$ be typed grounding, end-to-end target accuracy,
schema-qualified byte-exact replay, and the conditional frequency of the modal
parsed vector, respectively. The hardened ordinal release rule is
\begin{equation}
\operatorname{release}(\hat P_{\mathcal D_{\mathrm{pub}}})=
\begin{cases}
\hat P_{\mathcal D_{\mathrm{pub}}}, & \mathcal D_{\mathrm{pub}}\subseteq\mathcal D_{\mathrm{ctrl}}
        \land p\geq.95 \land c=1 \\
        & \land\ \min_{f,s}d_{fs}=1 \land g\geq.90 \land t\geq.90 \\
        & \land\ r\geq.95 \land m<.20,\\
\mathrm{WITHHOLD}, & \text{otherwise.}
\end{cases}
\label{eq:gate}
\end{equation}
The numeric thresholds and intervention concepts were fixed before the model
outputs. Replay semantics and evidence coverage received documented post hoc
conservative corrections that only add failures. Raw byte equality remains a
diagnostic: two invalid strings can be identical. The current controlled set
covers visual, information, and interaction repetition only; AI-estimated
reserve and clarity are not eligible for release in this paper.

\begin{figure}[t]
\centering
\includegraphics[width=\linewidth]{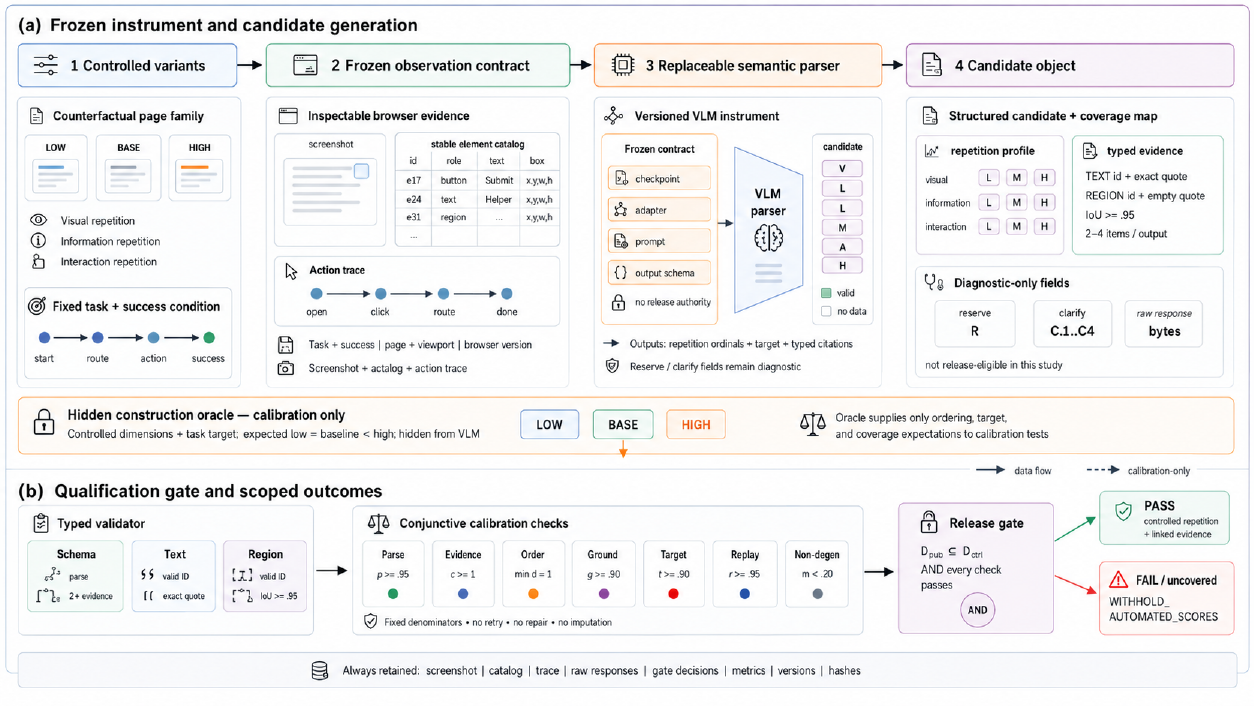}
\caption{Instrument architecture of \cora{}. A fixed browser contract produces
inspectable screenshots, catalogs, and traces. A replaceable, versioned VLM
proposes ordinal judgments and typed citations but has no direct release path.
A hidden construction oracle provides only controlled ordering, targets, and
coverage to fixed-denominator tests. The conjunctive gate releases an
evidence-linked repetition profile only for covered dimensions when every
check passes; otherwise it returns \withhold{}, while raw evidence and
diagnostic-only reserve and clarity remain available.}
\label{fig:workflow}
\end{figure}

Figure~\ref{fig:workflow} shows that no path leads directly from the VLM to a
released score. Qualification applies to the complete instrument configuration,
and the VLM cannot see the construction oracle. In this study, only controlled
repetition dimensions are release candidates; estimated reserve and clarity
remain diagnostic. If the semantic AI layer fails qualification, the raw
evidence and deterministic measurements are still retained.

\section{Experiments}

The empirical studies and implementation suites have different roles
(Table~\ref{tab:studies}). A parameterized rendered-page suite lets us control
the mechanism and instantiate known failure conditions. These experiments
therefore evaluate identifiability, auditability, reproducibility, and refusal
within a controlled benchmark; they do not measure human users.

\begin{table}[t]
\centering
\caption{Evidence roles. ``Domains'' in Studies C/D are task/content domains
sharing one HTML/CSS layout template.}
\label{tab:studies}
\small
\begin{tabularx}{\linewidth}{@{}P{0.12\linewidth}P{0.22\linewidth}P{0.18\linewidth}Y@{}}
\toprule
Component & Scope & Primary question & Claim boundary \\
\midrule
Study A & 1,555,200 primary episodes; five seeds total & Is reserve identifiable
and seed-robust at fixed load? & 7,776,000 planted-mechanism episodes \\
Study B & 77 Qwen calls, one page family & Can stability hide collapse? &
Visible labels and easy target \\
Study C & 336 Qwen calls, 42 pages & Does a generation-locked ordinal repair pass? & Six
domains, shared layout \\
Study D & 210 SmolVLM2 calls, same pages & Does the gate transfer unchanged? &
Two small local VLMs \\
Conformance & 504 citations; 48 gate panels & Does code implement the spec? &
32 general plus 16 coverage panels; no semantic validity \\
\bottomrule
\end{tabularx}
\end{table}

\subsection{Study A: exactly load-matched identifiability}

Study A uses six parameterized task configurations, three redundancy types,
load budgets 4/8/12, a $2\times2$ reserve-by-tax allocation, four perturbations,
three observation-noise levels, and 600 episodes per cell. All nine
type-by-budget strata have exactly equal scalar load and contain all four
reserve-tax compositions. Five replicates of 120 episodes yield 2,592 cells and
1,555,200 episodes.

Within family, type, load, perturbation, and noise strata, we regress retention
on matching reserve. High- and low-tax profiles are strictly paired. For
perturbed-success prediction, one family is left out at a time. Scalar
baselines receive no profile ID, reserve/tax labels, or component counts. A
500-tree random forest is the strongest scalar baseline; no hyperparameter
search is performed. Full equations and model inputs appear in
Appendix~\ref{app:study-a}.

A post-audit sensitivity follow-up reuses the primary seed and runs four
additional unchanged master seeds at the same 600 episodes per cell. The five
full runs contain 7,776,000 episodes. We report ranges and sample standard
deviations descriptively; the selected seeds are not treated as a population.

\subsection{Study B: qualification stress test}

Study B renders seven variants of one parameterized e-commerce page family: a
baseline and low/high visual, information, and interaction redundancy. The
local model is Qwen2.5-VL-3B-Instruct \cite{bai2025qwen25vl}, revision
\nolinkurl{66285546d2b821cf421d4f5eb2576359d3770cd3}. Screenshot-only,
evidence-only, and fused contracts use seeds 101, 202, and 303, plus two fused
temperature-zero replays per page, for 77 calls. The JSON example contains
literal zeros for load/reserve and hundreds for clarity. Condition labels and
an answer-revealing target summary are visible, making this a failure-discovery
study rather than a blinded qualification.

\subsection{Study C: generation-locked calibration repair}

Study C contains 42 rendered pages from travel, education, municipal service,
productivity, content discovery, and appointment booking. Each task/content
domain has baseline, visual-low/high, information-low/high, and
interaction-low/high variants. All six domains share one HTML/CSS card layout.
Rendered pages hide manipulation names, levels, variant IDs, and answer
summaries. The fused input still exposes deterministic structural counts such
as routes, helper text, decorative units, and confirmations, so the study is
blinded to condition names rather than to structure.

Two frozen arms evaluate every page with Qwen. \aabs{} retains Study B's
anchored 0--100 output contract. \cord{} requests low/medium/high/uncertain
ordinals, typed text-or-region evidence, and no numeric examples. Each arm uses
three stochastic seeds; \cord{} also has two temperature-zero replays per page:
$252$ stochastic calls plus $84$ replays, or $336$ total. Generation settings,
prompts, thresholds, and metric concepts were hash-locked before the relevant
outputs. Immutable-raw corrections later separated JSON syntax from schema and
grounding, fixed end-to-end denominators, enforced schema-qualified replay, and
added the already specified 2--4 evidence coverage check. The audit trail
records when each conservative correction occurred.

\subsection{Study D: frozen cross-model replication}

Study D runs \cord{} unchanged with SmolVLM2-2.2B-Instruct
\cite{marafioti2025smolvlm}, revision
\nolinkurl{482adb537c021c86670beed01cd58990d01e72e4}. Model-specific loading and
chat templating are the only adapter changes. The 42 pages receive three
stochastic seeds and two deterministic replays, for $126+84=210$ calls. The
checkpoint, prompt, stimuli, runner, thresholds, and original metric concepts
were locked before the first output. As in Study C, immutable-output analysis
later incorporated the documented conservative replay and evidence-coverage
corrections without changing any raw output. Neither prompt nor checkpoint was
substituted after execution began.

\subsection{Validator and gate conformance suites}

The typed grounding suite contains 504 balanced citations over the 42 Study C
pages: 252 valid and 252 invalid. It compares the legacy overlapping-quote rule
with the frozen typed-v2 rule. This tests specification conformance, not whether
model evidence is semantically relevant.

The end-to-end gate suite contains 16 valid and 16 invalid seven-page panels.
Each panel passes 21 raw strings through the real JSON extractor, ordinal
normalizer, typed validator, ordering and target aggregation, raw replay, and
degeneracy check, for 672 strings total. Faults include truncation, invalid
ordinals, reversed or shuffled order, invalid IDs, quote/region mismatch,
wrong target, unstable replay, constant vectors, and four composites. The
protocol was frozen before corpus construction and the deterministic pipeline
was run three times with hash-identical artifacts.

A separate post-audit evidence-cardinality suite contains eight valid and eight
invalid panels, or 336 raw strings. The invalid panels preserve correct ranks,
targets, citations, and replay while using empty, missing, non-list, sparse, or
overfull evidence on at least one output. It compares the prior pooled-grounding
gate with the hardened output-level 2--4 requirement. Its protocol was frozen
after discovery of the coverage gap but before corpus implementation and
execution.

\section{Results}

Study A tests whether the representation separates reserve from normal-use tax.
Studies B--D apply the same release logic to two model configurations, followed
by direct tests of the checker. All results below come from the pre-specified
controlled benchmark.

\subsection{\cora{} separates reserve from tax within the planted mechanism}

\begin{figure}[t]
\centering
\includegraphics[width=\linewidth]{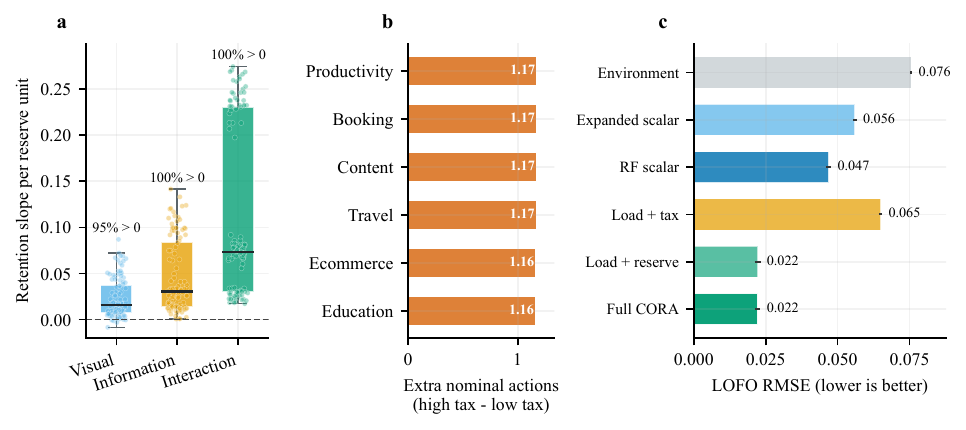}
\caption{Study A. (a) Retention slope per failure-domain-matched reserve unit;
annotations show fractions above zero. (b) Extra nominal actions for matched
high-tax profiles. (c) Leave-one-family-out perturbed-success RMSE; error bars
are descriptive 95\% $t$ dispersion intervals over six overlapping held-out
configurations.}
\label{fig:matched}
\end{figure}

Reserve had a positive retention slope in 319/324 analyzable strata (98.5\%;
Figure~\ref{fig:matched}a). Median within-stratum Spearman $\rho$ was 0.894,
and all six family median slopes were positive (exact two-sided sign test
$p=0.03125$ over six fixed configurations). Strictly matched high-tax profiles added 1.165 nominal actions on
average, with positive differences in 324/324 descriptive pairs and all six
family means.

Full \cora{} achieved LOFO RMSE 0.0222 and $R^2=0.9559$, versus RMSE 0.0469
and $R^2=0.8030$ for the random-forest scalar. The relative RMSE reduction was
52.8\%; all six overlapping paired folds favored \cora{} (descriptive
$p=0.03125$). Load plus reserve
was effectively identical to full \cora{}: mean RMSE differed by
$2.58\times10^{-6}$ and the paired interval crossed zero. Thus reserve predicts
failure survival inside the generator, whereas tax describes normal cost.

All five master seeds met the frozen follow-up criteria. Positive reserve-slope
strata ranged from 311/324 to 323/324; all six family medians and all 324 tax
pairs were positive in every run. Mean tax cost ranged from 1.158 to 1.165
actions. Full \cora{} RMSE ranged from 0.0204 to 0.0222, versus 0.0461--0.0470
for the random-forest scalar and 0.0549--0.0560 for the prospectively specified
expanded scalar. The follow-up therefore checks sensitivity to the random seed
while holding the generator and six families fixed.

\subsection{Qualification distinguishes repeatability from discrimination}

Study B shows the value of checking replay stability and construct
discrimination separately. Its outputs were highly repeatable and matched the
example, but they did not respond to controlled page changes.

Seventy-two of 77 Study B calls produced parseable JSON. Every parseable score
vector was
\begin{equation}
[0,0,0,0,100,100,100,100,100],
\end{equation}
and all five truncated outputs emitted the same model-supplied numeric prefix.
Thus 77/77 raw outputs copied the example's 0/100 pattern. Controlled ordering
was zero. Legacy catalog grounding was 93/135 (68.9\%), and the easy fused
target was correct in 20/21 scheduled runs. All 7/7 deterministic replay pairs
were byte-identical. \cora{} keeps these signals separate, so high stability and
easy-target success cannot substitute for construct discrimination.

\subsection{Qualification evaluates the complete output contract}

Study C changed the complete Qwen output contract. Ordering improved from
\aabs{} to \cord{}, while the other release criteria still failed. We report
the ordering gain, but no \cord{} score was eligible for release.

\begin{table}[t]
\centering
\caption{Model-contract qualification after documented conservative analysis
corrections. Schema, evidence count, and target use 126 scheduled stochastic
calls; strict order uses 108 family-seed relations; grounding uses emitted
items; qualified replay uses 42 pairs and requires schema-valid byte equality.
``N/A'' denotes no eligible denominator.}
\label{tab:qualification}
\small
\begin{tabular}{lrrrrrr}
\toprule
Instrument & Schema & Evidence & Strict & Ground & Target & Qualified \\
& parse & 2--4 & order & & & replay \\
\midrule
Qwen \aabs{} & 87.3 & 76.2 & 0.0 & 0.0 & 65.9 & N/A \\
Qwen \cord{} & 62.7 & 25.4 & 10.2 & 0.0 & 20.6 & 66.7 \\
SmolVLM2 \cord{} & 0.0 & 0.0 & 0.0 & N/A & 0.0 & 0.0 \\
\bottomrule
\end{tabular}
\end{table}

In Study C, \aabs{} parsed 110/126 calls but recovered 0/108 strict adjacent
relations and 0/162 pairwise relations. All 126 raw outputs contained the same
anchored numeric prefix; the 110 parsed vectors were identical. Target accuracy
was 83/126 (65.9\%), and no evidence item passed the arm's joint grounding
contract. Only 96/126 calls also satisfied the specified 2--4 evidence count.

\cord{} parsed 79/126 calls. Strict ordering rose to 11/108, or 10.2\%
(family-bootstrap 95\% CI 6.5--13.9), and pairwise accuracy was 17/162, or
10.5\% (CI 6.8--14.8). The family-level \cord{} minus \aabs{} ordering
difference was +10.2 percentage points; all six fixed family means were positive
(exact directional summary $p=0.03125$). The two arms changed numeric versus
ordinal scales, schema, evidence format, and examples together, so this result
shows sensitivity to the complete output contract rather than isolating numeric
anchors. Typed grounding was 0/358, only 32/126 calls satisfied schema plus the
2--4 evidence count, target accuracy fell to 26/126 (20.6\%), and the modal
parsed vector occurred in 16/79 cases (20.25\%), just above the strict
non-degeneracy limit.

All 42 Qwen \cord{} raw pairs were byte-identical, yet only 28 pairs were
schema-valid on both calls. Protocol-qualified replay was therefore 28/42
(66.7\%), not 42/42. Parse, evidence coverage, ordering, grounding, target,
qualified replay, and non-degeneracy checks all failed, so the result was
\withhold{}.

\subsection{The same qualification gate supports cross-model comparison}

Study D tests whether the qualification logic can remain fixed when the model
changes. The same ordinal contract and release gate were applied to SmolVLM2
without model-specific prompt or threshold changes, and the unchanged gate
identified a different unsupported output pattern.

SmolVLM2 produced syntactically extractable JSON in 82/126 stochastic calls but
no schema-valid ordinal object. The 82 JSON objects copied the deterministic
observation trace rather than the requested measurement; the other 44 outputs
were unbalanced. Consequently, strict and pairwise ordering and target accuracy
were all zero, while grounding and vector-degeneracy denominators were absent.
All 42 raw pairs were again byte-identical, but protocol-qualified replay was
0/42 because neither member of any pair was schema-valid. Relative to Qwen \cord{},
SmolVLM2 ordering was 10.2 percentage points lower (family-bootstrap CI
$-13.9$ to $-6.5$); all six family differences were negative (exact sign test
$p=0.03125$ over six fixed domains). The same dimension-scoped gate withheld
the second instrument.

The unchanged procedure made the model-specific violations directly comparable:
Qwen produced partly parseable but poorly grounded ordinal judgments, whereas
SmolVLM2 echoed traces without schema compliance. This replication does not
measure agreement between models about website quality.

\subsection{Executable conformance tests verify the release path}

The conformance suites test the typed validator and the complete release path
independently of model quality. The separate evidence-cardinality suite verifies
the output-level requirement added after the coverage gap was identified.

\begin{figure}[t]
\centering
\includegraphics[width=\linewidth]{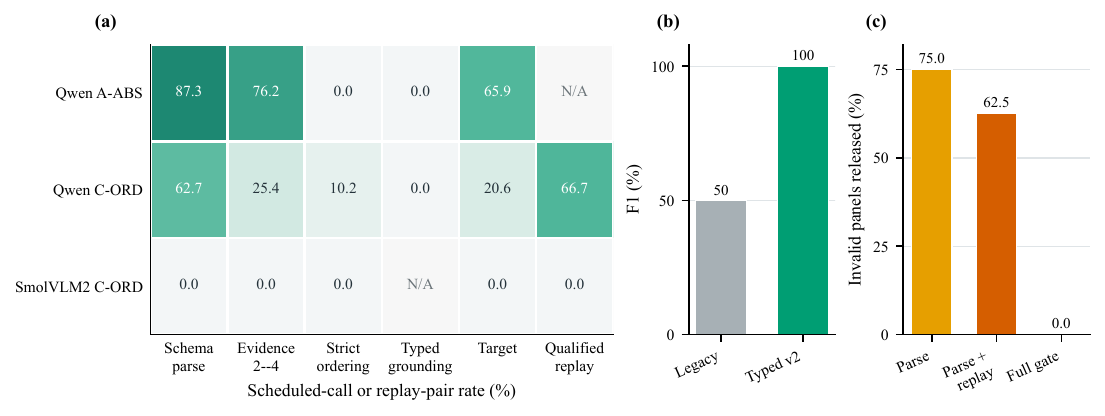}
\caption{Calibration and implementation results. (a) Qualification metrics
for the three model-contract instruments, with distinct denominators: scheduled
calls, family-seed relations, emitted evidence items, or replay pairs. Evidence
2--4 is scheduled-call compliance; qualified replay requires schema-valid byte
identity. Both models separately had 42/42 raw byte-equal pairs. (b) The
typed-v2 validator versus the legacy
rule on 504 balanced specification cases. (c) Invalid panels released by
incomplete and full gates on the 32-panel, 672-string conformance suite. Panels
(b,c) measure specification conformance, not model grounding or real-world
error rates.}
\label{fig:calibration}
\end{figure}

On the citation suite, typed v2 achieved F1 1.00 with neither false accepts nor
false rejects. The legacy validator had F1 0.50, with 126 false accepts and 126
false rejects, and the improvement was identical across all six page families.
This establishes implementation conformance on the frozen citations. Study C
grounding remains 0/358.

On the end-to-end gate suite, the full gate released all 16 valid panels and
withheld all 16 invalid panels. By comparison, parse-only released 12/16 invalid
panels (75.0\%), and parse plus raw replay released 10/16 (62.5\%). The 0/16
count describes coverage of these prospectively enumerated faults; no
production error rate is inferred from it.

The evidence-cardinality hardening suite tested a gap not covered by the first
gate fixtures. The prior full gate released all 8/8 invalid panels because
their emitted citations were individually valid and pooled grounding remained
100\%. Requiring every scheduled output to contain 2--4 evidence items released
all 8 valid panels and 0/8 invalid panels. This test covers the enumerated
cardinality faults and no broader fault family.

\section{Discussion}

\subsection{What repeated runs can establish}

Running the same model configuration again is straightforward. The harder
question is whether its output can be released as a measurement. \cora{} fixes
the observation and intervention, preserves the evidence, and keeps the
denominator and release rule stable. We call this property \emph{procedural
objectivity}: another researcher can inspect what the model saw, which checks
failed, and why a score was released or withheld. The term applies to the audit
procedure. It does not establish that the model is unbiased or correct; human
agreement still requires a separate study.

A withheld run is still informative. The audit retains the screenshot, page
catalog, task trace, raw response, and failed checks. In the two model studies,
these records showed why stable outputs did not qualify as measurements of
website quality.

\subsection{Interpreting the contract comparison}

The Qwen contract comparison illustrates the distinction. Ordering increased
in every fixed family mean, while schema validity and target accuracy
deteriorated; grounding stayed at zero, and the modal-vector threshold failed.
Because the scale, schema, evidence format, and examples changed together, the
study cannot attribute the ordering gain uniquely to numeric anchors. We
therefore report it as an ordering result. It does not override the other
failures in the instrument.

Under the same ordinal prompt, the two models violated the contract differently.
Qwen returned partly parseable ordinal judgments with poor grounding. SmolVLM2
copied observation traces. Separate failure records make that difference
visible.

\subsection{Requirements for broader use}

Before this workflow could serve as a standard, it would need fixed tasks and
rendering, plus inspectable text or region evidence for every output. Each
reported dimension would need its own controlled intervention and a coverage
map that blocks release elsewhere. Parsing, ordering, grounding, target use,
replay, and degeneracy would remain separate checks. External calibration
across sites, experts, users, languages, devices, and accessibility conditions
is still missing. The prototype implements internal checks for three repetition
dimensions; AI-estimated reserve and clarity remain diagnostic.

The load-tax-reserve representation gives designers more specific questions to
test. Visual repetition without an independent failure domain is a candidate
for removal. Information repetition should be tested for cue independence
rather than counted as repeated text. An alternate route should count as
reserve only if it survives a different failure. Confirmation or decoy steps
can add tax without reserve. These hypotheses arise from the mechanistic
benchmark and still require tests on real interfaces.

\section{Limitations}

The current evidence addresses the internal validity of the planted mechanism
and the conformance of the evidence validator and release code. The experiments
used neither human participants nor production sites, so they provide no
inter-rater reliability, expert agreement, perceived clarity, task time,
preference, or observed recovery behavior. AI-human comparisons require a
separate study.

Study A is an identifiability experiment in one planted mechanism. Its six
families are parameter settings rather than production websites or distinct
graph topologies. The episode count reduces Monte Carlo noise, and the five-seed
follow-up measures seed sensitivity within the same generator. Family-level
summaries still contain only six fixed units: $p=0.03125$ is the minimum
two-sided sign-test value when all six directions agree. Because no website
population was sampled, the overlapping-fold intervals describe dispersion
rather than population uncertainty.

Study B exposes labels and a visible answer summary. Study C removes those
fields, but its six task/content domains share one layout and its trace reveals
deterministic structural counts. Studies C and D measure contract response under
controlled benchmark variation rather than open-world visual understanding.
The model sample comprises two small local VLMs evaluated on English desktop
pages with one viewport and one GPU environment.

The oracle independently manipulates visual, information, and interaction
repetition. It does not manipulate AI-estimated reserve or the four clarity
fields, which remain diagnostic. Validator F1 1.00 and the gate's 0/16 false
releases apply to frozen specification fixtures. They neither establish
semantic relevance nor exhaust arbitrary parser attacks, and they provide no
production error rate. Model grounding still fails even though the validator
passes these fixtures. The 0.90 and 0.95 thresholds are engineering choices
without human calibration. We designed the evidence-cardinality suite after
finding that vulnerability, so its 0/8 result covers the enumerated faults
rather than unknown coverage failures.

The Study C protocol text originally stated that layout was randomized by
family, but the rendered artifact uses one shared layout template. The
immutable correction record and this paper follow the artifact-grounded
description. We corrected schema/grounding separation during execution, then
corrected replay semantics and evidence coverage after execution. The
corrections were conservative and recorded in the audit trail; raw model outputs
were unchanged. Because of an import-path packaging defect, the analyzer
requires the repository root on \texttt{PYTHONPATH} (or module invocation).

The evaluation excludes multilingual, mobile, accessibility-specific, and
unconstrained computer-use-agent settings. Redundancy may have different value
for screen-reader, keyboard, low-vision, motor, or cognitive-access tasks. Each
setting requires explicit failure domains and outcomes.

\section{Ethics, data governance, and AI assistance}

The studies use programmatically rendered benchmark pages, generated task
trajectories, and local open model checkpoints, and collect no personal data.
Human or production studies would require appropriate consent, no retention of
personal form content, masking of typed payloads, and treatment of cursor and
interaction traces as research data.
Reports should separate deterministic observations from model inference and
state the calibration status.

We used generative AI tools during code development, experiment orchestration,
analysis checks, and manuscript drafting. Numerical claims were checked against
machine-readable artifacts or deterministic validators. Tool assistance is not
treated as authorship or as independent evidence.

\section{Reproducibility and data availability}

The local artifact contains frozen protocols, lock files, raw model outputs,
corrected analyzers, manifests, hashes, compact metrics, figure scripts,
citation verification records, and end-to-end conformance tests. Study A's
eight primary artifacts were regenerated byte-for-byte at the same seed, and
four additional full master seeds are stored with manifests and aggregate
metrics. Study C preserves all 336 raw calls; post hoc conservative replay and
evidence-coverage errata bind old and new analyzer hashes to immutable raw
outputs. Study D records the complete checkpoint revision and model-file
hashes. Both gate suites produced identical deterministic artifacts in three
executions.

No persistent public repository URL has been assigned at the time of this
preprint. A public code/data release is still needed for external reproduction;
the arXiv source package contains the manuscript and vector figures but not the
multi-gigabyte model checkpoints or full episode table.

\section{Conclusion}

\cora{} treats website-redundancy judgment as a measurement problem that can be
tested. In Study A, the separation of load, tax, and reserve recovered the
planted behavior across five master seeds. The model studies produced repeatable
outputs, but neither VLM passed the release checks. Qwen showed a small ordering
gain and failed grounding and coverage. SmolVLM2 did not return the requested
schema. The checker recorded both failure patterns and withheld both scores.
After the audit exposed an evidence-cardinality gap, a separate fault suite
tested the hardened rule.

At present, \cora{} is an executable measurement proposal for external testing.
Its main contribution is a reproducible account of why the procedure released
or withheld a score, including the page evidence and failed checks. It will
require independent sites and human judgments across languages, devices, and
access needs before it can serve as a general standard.

\bibliographystyle{plainnat}
\bibliography{references}

\clearpage
\appendix

\section{Study A mechanistic specification}
\label{app:study-a}

For budget $B$, low/high reserve receive $B/4$ and $B/2$ units; low/high tax
receive 0 and $B/4$ units; neutral units fill the remainder. Visual profiles
allocate units among landmarks, decoration, and taxing friction; information
profiles use independent cues, duplicates, and semantic noise; interaction
profiles use routes, decoys, and confirmations. Other dimensions stay at
medium settings.

Given available signal $s$, routes $r$, family ambiguity $a$, observation
noise $n$, and distraction burden $d$, the mechanistic correct-action
probability is
\begin{equation}
\begin{aligned}
p_{\mathrm{correct}}
&=\operatorname{clip}_{[0.10,0.97]}\,\sigma\Bigl(
2.35-2.05a-2.25n \\
&\qquad +0.64\log(1+s)+0.28\log(1+r)
-0.38\log(1+d)\Bigr).
\end{aligned}
\end{equation}
Matching structural-loss probability is 0.45. Tax units produce stochastic
nominal actions with probability $\min(0.85,0.50+0.35n)$. The mechanism is
transparent by design and should not be interpreted as a model of human
behavior.

\begin{table}[htbp]
\centering
\caption{Study A leave-one-family-out perturbed-success prediction.}
\small
\begin{tabular}{lrr}
\toprule
Representation & RMSE $\downarrow$ & $R^2$ $\uparrow$ \\
\midrule
Environment only & 0.0756 & 0.4879 \\
Scalar load & 0.0649 & 0.6227 \\
Expanded scalar & 0.0560 & 0.7195 \\
Random-forest scalar & 0.0469 & 0.8030 \\
Load + tax & 0.0649 & 0.6227 \\
Load + reserve & 0.0222 & 0.9559 \\
Full \cora{} & \textbf{0.0222} & \textbf{0.9559} \\
\bottomrule
\end{tabular}
\end{table}

\section{Study B prompt diagnostics}

The exact match between Study B scores and the JSON example is consistent with
numeric-template anchoring but does not prove causality. Two post-hoc,
temperature-zero seven-page diagnostics were stored separately. Removing the
example produced 0/7 usable score objects; the model mainly echoed the trace or
truncated. Replacing numeric examples with uppercase string placeholders also
produced 0/7 numeric-schema-compliant outputs. These diagnostics motivated the
prospective \cord{} arm but are not pooled with any confirmatory result.

Study B's 68.9\% grounding uses a legacy ID-plus-overlapping-quote rule. It is
not directly comparable to Study C's typed contract. Thirty-six invalid Study B
items named valid textless visual elements, which motivated the region
modality. The typed-v2 conformance result validates that new rule on fixtures;
it does not retrospectively recode model evidence as semantically relevant.

\section{Minimum external-validation study}

A next study should use independently authored layouts from multiple domains,
languages, and device classes. Human participants and at least two versioned
computer-use agents should complete nominal, cue-loss, and route-loss tasks.
Separate controlled interventions should vary reserve and each clarity
dimension before those outputs become release-eligible.
Primary outcomes should include success, recovery success, action count, task
time, time to first relevant action, and post-failure recovery steps. Experts
should annotate evidence relevance and failure-domain independence without
seeing model scores. Subjective redundancy, clarity, and aesthetics should
remain separate outcomes rather than labels for one scalar score. The release
thresholds should be preregistered or calibrated on a development set and
evaluated once on held-out sites.

\end{document}